\documentclass[aps,prb,twocolumn,longbibliography,superscriptaddress]{revtex4-2}%
\usepackage{amsfonts}
\usepackage{mathrsfs}
\usepackage{amsmath}% needed for subequations
\usepackage{color}
\usepackage{physics}
\usepackage{soul}
\usepackage{siunitx} % si unit
\usepackage{graphicx}
\usepackage{bm}% bold maths
\usepackage{amssymb}
\usepackage{float}
\usepackage[normalem]{ulem}
\usepackage[colorlinks=true, letterpaper=true, pdfstartview=FitV, linkcolor=blue, citecolor=blue, urlcolor=blue]{hyperref}

\makeatother
\begin{document}

\title{Nonlinear current response of two-dimensional systems under in-plane magnetic field}

\author{Yue-Xin Huang}
\affiliation{Research Laboratory for Quantum Materials, Singapore University of Technology and Design, Singapore 487372, Singapore}
\affiliation{Department of Physics, City University of Hong Kong, Kowloon, Hong Kong SAR}

\author{Yang Wang}
\affiliation{Department of Physics and Astronomy, University of Delaware, Newark, Delaware 19716, USA}

\author{Hui Wang}
\affiliation{Research Laboratory for Quantum Materials, Singapore University of Technology and Design, Singapore 487372, Singapore}

\author{Cong Xiao}
\email{congxiao@hku.hk}
\affiliation{Department of Physics, The University of Hong Kong, Hong Kong, China}
\affiliation{HKU-UCAS Joint Institute of Theoretical and Computational Physics at Hong Kong, China}

\author{Xiao Li}
\email{xiao.li@cityu.edu.hk}
\affiliation{Department of Physics, City University of Hong Kong, Kowloon, Hong Kong SAR}

\author{Shengyuan A. Yang}
%\email{shengyuan_yang@sutd.edu.sg}
\affiliation{Research Laboratory for Quantum Materials, Singapore University of Technology and Design, Singapore 487372, Singapore}

\begin{abstract}
We theoretically investigate the nonlinear response current of a two-dimensional system under an in-plane magnetic field. Based on the extended semiclassical theory,
we develop a unified theory including both longitudinal and transverse currents and classify contributions according to their scaling with the relaxation time. Besides time-reversal-even contributions, we reveal a previously unknown time-reversal-odd contribution to the Hall current, which occurs in magnetic systems, exhibits band geometric origin, and is linear in relaxation time.
We show that the different contributions exhibit different symmetry characters, especially in their angular dependence
on the field orientation, which can be used to distinguish them in experiment. The theory is explicitly demonstrated in the study of the Rashba model.
Our work presents a deepened understanding of nonlinear planar transport, proposes approaches to distinguish different contributions, and sheds light on possible routes to enhance the effect in practice.
\end{abstract}
\maketitle

\section{Introduction}
Electronic transport properties are a central topic in condensed matter physics. They are widely used as basic tools to
characterize specific materials, to distinguish different states of matter, and to extract information about
underlying microscopic features. They also form the basis for most electronic device applications. Recent theoretical and experimental works have extended the study to various nonlinear transport effects~\cite{pacchioni_hall_2019,du_nonlinear_2021}. Importantly, it was shown that these nonlinear effects may manifest intriguing band structure quantities not accessible in linear responses, such as the Berry curvature dipole~\cite{sodemann_quantum_2015,kang_nonlinear_2019,ma_observation_2019}, Berry connection polarizability (BCP)~\cite{gao_field_2014,lai_third-order_2021,liu_berry_2022}, anomalous spin/orbital polarizability~\cite{wang_theory_2022,xiao_time-reversal-even_2023}, and etc. Because of this and their potential application in nonlinear devices, such effects have been attracting great interest~\cite{facio_strongly_2018,du_band_2018,battilomo_berry_2019,matsyshyn_nonlinear_2019,he_quantum_2021,tiwari_giant_2021,he_graphene_2022,sinha_berry_2022,duan_giant_2022}.

One of the nonlinear transport effects that has been actively explored is the nonlinear electric current response
in the presence of an in-plane magnetic field, namely, the response current scales as $j\sim E^2 B$,
and the $B$ field lies in the two-dimensional (2D) plane (the transport plane) of the response current and the driving $E$ field. The response current has a longitudinal component (i.e., component along the driving $E$ field), which is connected to effects known as bilinear magnetoresistance or unidirectional magnetoresistance~\cite{rikken_observation_2002,ideue_bulk_2017,he_bilinear_2018,choe_gate-tunable_2019,guillet_observation_2020,li_nonreciprocal_2021,wang_large_2022}. It also has a transverse component, which has been probed in 2D systems such as surface of a topological insulator and some 2D materials~\cite{he_nonlinear_2019,wang_observation_2022}, and in topological semimetal CoSi~\cite{esin_nonlinear_2021}. Several previous theories proposed extrinsic mechanisms for this effect, with resulting current $\sim\tau^2$~\cite{he_nonlinear_2019,zheng_origin_2020,battilomo_anomalous_2021,rao_theory_2021,dantas_determination_2022,marinescu_magnetochiral_2023}, where $\tau$ is the relaxation time. In a recent work~\cite{huang_intrinsic_2023}, we showed that there actually exists an intrinsic contribution to this nonlinear current which
gives a dissipationless (Hall) response, i.e., the resulting current is independent of scattering ($\sim \tau^0$) and satisfies $j_a E_a=0$ ($a$ labels Cartesian components and the Einstein summation convention is assumed in this paper).

The current work serves three purposes. First, we develop a unified theory for this nonlinear effect in 2D systems, which captures all previously proposed mechanisms. Our theory is based on the extended semiclassical theory framework~\cite{gao_field_2014,gao_geometrical_2015,gao_semiclassical_2019} and at the level of relaxation time approximation. The obtained formulas have the advantage of being readily applicable to model calculations or combined with first-principles approaches. Second, we predict that in 2D systems with broken time reversal symmetry, there is a previously unknown contribution which scales linearly in $\tau$. Interestingly, we find that this contribution encodes information about the anomalous spin polarizability (ASP), which is an important geometric quantity of band structure for not only nonlinear transport but also magnetoelectric response and spin-orbit torque. Third, we demonstrate the application of our theory to the Rashba model, which highlights the different behaviors of different contributions and offers guidance for separating them in experimental studies. By accomplishing these three targets, the present work will not only deepen our understanding of this intriguing nonlinear effect, but also be useful for subsequent theory development, for predicting suitable material platforms, and for interpreting experimental results.

\section{The General theory}

\subsection{Basic setup}
We consider a 2D system in the $x$-$y$ plane. As mentioned in the introduction, we are concerned with the setup in which both the applied $E$ field and $B$ field lie in the 2D ($x$-$y$) plane, and we consider the dc limit. Experimentally, the nonlinear signal is usually extracted by lock-in technique with a low-frequency modulation on the driving $E$ field~\cite{ma_observation_2019,kang_nonlinear_2019,lai_third-order_2021}. The dc limit means the modulation frequency always stays in the $\omega\tau\ll 1$ regime.

The type of nonlinear current we are looking for can be expressed in terms of a nonlinear conductivity tensor $\chi$ as
\begin{eqnarray}
	j_a= \chi_{abcd} E_b E_c B_d,
	\label{eq-typecurrent}
\end{eqnarray}
where the subscripts $a,b,c,d\in \{x,y\}$ according to our setup.
Regarding its symmetry character, $\chi_{abcd}$ is a fourth-rank axial tensor. Therefore, it vanishes if the system has an inversion center or a horizontal mirror plane $M_z$. Describing the transport process, $\chi$ has both time-reversal ($\mathcal{T}$) even and odd parts, which are distinguished according to their parities under $\mathcal{T}$~\cite{freimuth_spin-orbit_2014,zelezny_spin-orbit_2017,manchon_current-induced_2019,xiao_time-reversal-even_2023}. Particularly, the $\mathcal{T}$-odd part of $\chi$ can only appear in magnetic systems, whereas the $\mathcal{T}$-even part is allowed in both nonmagnetic and magnetic systems. Moreover, in magnets the two parts have different transformation properties under any primed operation which is a combination of $\mathcal{T}$ and a spatial operation.

To investigate this nonlinear current, we adopt the approach of extended semiclassical theory, which includes higher-order field corrections not present in usual semiclassical theory. This theory has found great success in describing many nonlinear response properties~\cite{lai_third-order_2021,liu_intrinsic_2021,wang_intrinsic_2021,xiao_intrinsic_2022,xiang_third-order_2023}.

In this theory, the semiclassical equations of motion for a Bloch electron wavepacket which is centered at $(\bm r,\bm k)$ in phase space and has band index $n$ take the form of (take $e=\hbar=1$):
\begin{align}
    \dot{\bm r}&= \pdv{\tilde\varepsilon_n}{\bm k}- \dot{\bm k}\times \tilde{\boldsymbol\Omega}_n,
    \\
    \dot{\bm k}&= -\bm E-\dot{\bm r}\times \bm B.
    \label{eq-dotrk}
\end{align}
These equations have the same structure as the Chang-Sundaram-Niu equation of the conventional semiclassical theory~\cite{chang_berry_1995,sundaram_wave-packet_1999,xiao2010}, except that here the band energy $\tilde\varepsilon_n$ and the Berry curvature $\tilde{\boldsymbol\Omega}_n$ include field corrections~\cite{gao_field_2014}, as highlighted by the tilde in these symbols (and similar quantities without tilde are defined in terms of unperturbed band structure). Their specific expressions will be discussed in a while (and can be found in Ref.~\cite{gao_field_2014}). It should be noted that in our setup, the $B$ field is in the plane of the 2D system, so it will not give the Lorentz force term in equation of motion (\ref{eq-dotrk}). Nevertheless, it modifies the band structure hence enters $\tilde\varepsilon_n$ and $\tilde{\boldsymbol\Omega}_n$. Since the orbital motion is confined in the 2D plane, the in-plane orbital magnetic moment is suppressed, and the applied $B$ field will mainly couple with the spin magnetic moment of electrons in the systems we consider.

To calculate the nonlinear current, the equation of motion is combined with the Boltzmann equation within the relaxation time approximation. For a homogeneous system, we have
\begin{equation}
  \dot{\bm k}\cdot \nabla_{\bm k}f= -\frac{f-f_0}{\tau}
\end{equation}
where $f$ is the distribution function, and
$f_0$ is the equilibrium Fermi-Dirac distribution. In the semiclassical regime,
the solution to this equation can be formally written as
\begin{eqnarray}
    f=\sum_{\eta=0}^\infty (-\tau \dot{\bm k}\cdot \nabla_{\bm k})^\eta f_0(\tilde\varepsilon).
    \label{eq-fdis}
\end{eqnarray}

With the equation of motion and the solution of distribution function, one can calculate the charge current as
\begin{eqnarray}
    \bm j= -\int[\dd{\bm k}]\mathcal D(\bm k)\ \dot{\bm r}f ,
    \label{eq-current}
\end{eqnarray}
where $[\dd{\bm k}]$ is a shorthand notation for $\sum_n \dd{\bm k}/(2\pi)^2$ in a 2D system, and $\mathcal D(\bm k)$ is a correction factor for the phase space density of states~\cite{xiao_berry_2005} and it is $1$ for our present setup.
The desired nonlinear current $j$ can be extracted from Eq.~(\ref{eq-current}) by collecting the contributing terms that scale as $\sim E^2B$.
In the following, we shall group the contributions to $j$ (or $\chi$) according to their scaling relation with $\tau$. In the current theoretical framework, this leads to three types of currents.

\subsection{$\tau^0$-scaling current}
Let us first consider the contribution $\sim\tau^0$. This current, denoted as $j^{(0)}$, is independent of scattering, so it is called the \emph{intrinsic} contribution. The corresponding response tensor $\chi^{(0)}$ is entirely determined by the band structure, manifesting its significance as an intrinsic material property. The theory of this intrinsic nonlinear current has been developed in our recent work Ref.~\cite{huang_intrinsic_2023}. Here, we just briefly summarized the result.

At $\tau^0$ order, only the equilibrium Fermi-Dirac distribution enters into the current expression (\ref{eq-current}).
Combined with the equation of motion, one can easily see that the $\tau^0$ transport current only comes from the following
\begin{eqnarray}
    \bm j^{(0)}= - \int[\dd{\bm k}] (\bm E\times \tilde{\boldsymbol\Omega}_n) f_0(\tilde\varepsilon_n).
    \label{eq-j0}
\end{eqnarray}
Here, it should be understood that the desired current $j^{(0)}$ is to take the terms on the right hand side that scale as $\sim E^2B$. There is already a factor of $E$ in  the parenthesis. Note that the $E$ field correction to $\tilde\varepsilon_n$ is at least of second order. Hence, the remaining $E$ factor has to come from its correction to the Berry curvature $\tilde{\boldsymbol\Omega}_n$. This $E$ field correction takes the form of ${\bm\Omega}^E=\nabla_{\bm k}\times \bm{ \mathcal{A}}^E$, where $\mathcal{A}^E_a=G_{ab}E_b$ is the $E$-field induced Berry connection (or positional shift of wavepacket center)
and $G_{ab}$  is known as the BCP tensor which is a gauge invariant quantity~\cite{gao_field_2014}. For a band with index $n$, it can be expressed as
\begin{eqnarray}
    G_{ab}^n(\bm k)= 2\mathrm{Re} \sum_{m\neq n}\frac{v_a^{nm}v_{b}^{mn}}{(\varepsilon_n-\varepsilon_m)^3},
    \label{eq-Gabn}
\end{eqnarray}
where the $v$'s are the velocity matrix elements $v_a^{nm}=\langle u_n|\hat{v}_a|u_m\rangle$ with $|u_n\rangle$ the cell-periodic Bloch eigenstate. Finally, the remaining $B$ factor dependence comes from its correction to the band structure. (As mentioned before, the in-plane $B$ field does not produce a Lorentz force.) Here, its correction enters $\tilde\varepsilon_n$ and also
translates into a correction to $G_{ab}$. Explicitly, we have
\begin{equation}
  \tilde\varepsilon_n(\bm k)=\varepsilon_n(\bm k)-\boldsymbol{\mathcal{M}}^n\cdot \bm B,\label{11}
\end{equation}
where $\boldsymbol{\mathcal{M}}^{mn}=-g\mu_B{\bm s}^{mn}$ is the spin magnetic moment matrix element, with $g$ the $g$-factor, $\mu_B$ the Bohr magneton, and $\boldsymbol{s}^{mn}$ the spin matrix element; and $\boldsymbol{\mathcal{M}}^{n}\equiv \boldsymbol{\mathcal{M}}^{nn}$ denotes the diagonal element. The $B$ field correction to the BCP tensor $G_{ab}$ can be expressed as $\Lambda_{abc}B_c$, with
\begin{widetext}
\begin{align}
    \Lambda_{abc}^{n}(\bm k) =
    2\mathrm{Re}\sum_{m\neq n}
    \bigg[
        \frac {3v_{a}^{nm}v_{b}^{mn}\left(  \mathcal{M}_{c}^{n}-\mathcal{M}_{c}^{m}\right) }{(\varepsilon_{n}-\varepsilon_{m})^{4}}
        -\sum_{\ell\neq n}
        \frac{\left( v_{a}^{\ell m}v_{b}^{mn}+v_{b}^{\ell m}v_{a}^{mn}\right)  \mathcal{M} _{c}^{n\ell}}
        {\left(  \varepsilon_{n}-\varepsilon_{\ell}\right)  (\varepsilon _{n}-\varepsilon_{m})^{3}}
        -\sum_{\ell\neq m}
        \frac{\left( v_{a}^{\ell n}v_{b}^{nm}+v_{b}^{\ell n}v_{a}^{nm}\right)  \mathcal{M} _{c}^{m\ell}}
        {\left(  \varepsilon_{m}-\varepsilon_{\ell}\right)  (\varepsilon _{n}-\varepsilon_{m})^{3}}
    \bigg]
\end{align}
\end{widetext}
being interpreted as the spin susceptibility of BCP.

Collecting all the above relevant terms, we obtain the expression for the intrinsic nonlinear conductivity tensor
\begin{eqnarray}
    \chi_{abcd}^{(0)}= -\int [\dd{\bm k}] \Theta_{abcd}^n(\bm k) f_0'(\varepsilon _{n}),
    \label{eq-chiabcd0}
\end{eqnarray}
with
\begin{equation}
    \label{lambda}
    \Theta_{abcd}^{n}=
    (v_{a}^{n}\Lambda_{bcd}^{n}-v_{b}^{n}\Lambda_{acd}^{n})
    +\left( \partial_{a}G_{bc}^{n}-\partial_{b}G_{ac}^{n}\right)\mathcal{M}_{d}^{n},
\end{equation}
where $\partial_a$ is a shorthand notation for $\partial/\partial {k_a}$.

We have a few remarks before proceeding. First, in the final expression Eqs.~(\ref{eq-chiabcd0}, \ref{lambda}) for $\chi_{abcd}^{(0)}$, every quantity is expressed in terms of the intrinsic (unperturbed) band structure of the system, clearly manifesting its intrinsic character. Second, the $f_0'$ factor in (\ref{eq-chiabcd0}) shows that $\chi_{abcd}^{(0)}$ is a Fermi surface property, which is consistent with the general requirement of Fermi liquid theory~\cite{phillips_advanced_2012}. Third, the intrinsic currents found here is dissipationless, which can be easily seen from Eq.~(\ref{eq-j0}) that $j_a^{(0)} E_a=0$. It can also be observed that $\chi_{abcd}^{(0)}$ written in the form of (\ref{lambda}) is antisymmetric in its first two indices. Note that in Eqs.~(\ref{eq-chiabcd0}, \ref{lambda}), we have not symmetrized the two mid indices $b$ and $c$, which can always be done at the final step if needed. Fourth, the intrinsic current quantified by $\chi_{abcd}^{(0)}$ is $\mathcal T$-even, as can be verified by the
$\mathcal T$-invariance of Eq.~(\ref{eq-chiabcd0}). Finally, in this planar setup, since the effect of $B$ field is only to correct the band structure (through a Zeeman type coupling), the nonlinear current $j^{(0)}$ may be regarded as effectively an intrinsic second-order anomalous Hall effect in Ref.~\cite{liu_intrinsic_2021} applied to the $B$-field corrected band structure.

\subsection{$\tau^1$-scaling current}
As we have mentioned, when the time-reversal symmetry $\mathcal T$ is broken, the $E^2B$ current response allows a $\mathcal T$-odd contribution. Under the relaxation time approximation, this contribution has the form of $j^{(1)}\sim \tau$. Such a Hall current has up to now only been discussed in three dimensional magnetic Weyl semimetals based on the chiral anomaly mechanism~\cite{li_nonlinear_2021,nandy_chiral_2021}, but has not been proposed for 2D systems before.

In our extended semiclassical formalism, $j^{(1)}$ must involve the nonequilibrium distribution in (\ref{eq-fdis}) at linear order in $\tau$. This automatically gives a factor of $E$. To obtain a current in (\ref{eq-current}) at the second order of $E$ field, we only need to keep terms in $\dot{\bm r}$ that is linear in $E$, resulting in
\begin{equation}
  \bm j^{(1)}=\int [\dd{\bm k}](\bm E\times \tilde{\boldsymbol\Omega}_n) (-\tau \bm E\cdot \partial_{\bm k})f_0(\tilde{\varepsilon}_n).\label{15}
\end{equation}
On the right hand side only terms $\sim E^2B$ should be retained. Since there are two $E$ factors in the integrand, we only need to keep the linear in $B$ correction to $\tilde{\boldsymbol\Omega}_n$ and $\tilde{\varepsilon}_n$. The correction for $\tilde{\varepsilon}_n$ is the same as in Eq.~(\ref{11}). Meanwhile, the $B$-field correction to $\tilde{\boldsymbol\Omega}_n$ can be expressed as
${\bm\Omega}^B=\nabla_{\bm k}\times \bm{ \mathcal{A}}^B$, where $\mathcal{A}^B_a=F_{ab} B_b$ and $F_{ab}$ is also a gauge-invariant quantity, known as the ASP tensor~\cite{wang_theory_2022,xiao_time-reversal-even_2023}, which describes the spin magnetoelectric coupling of Bloch electrons, i.e., a positional shift induced by $B$ field and a anomalous spin magnetic moment induced by $E$ field. For a band with index $n$, ASP can be expressed as
\begin{eqnarray}
    F_{ab}^n(\bm k)= - 2\text{Im}\sum_{m\ne n} \frac{v_a^{nm}\mathcal{M}_{b}^{mn}}{(\varepsilon_n-\varepsilon_m)^2}
    \label{eq-Fab},
\end{eqnarray}

Substituting these corrections into Eq.~(\ref{15}), we obtain the conductivity tensor $\chi_{abcd}^{(1)}$ corresponding to current $j^{(1)}$:
\begin{eqnarray}
    \chi_{abcd}^{(1)}=- \tau\int[\dd{\bm k}]
    \Gamma^n_{abcd}(\bm k) f'_0(\varepsilon_n),
    \label{eq-chi1}
\end{eqnarray}
where we define the quantity
\begin{eqnarray}
    \Gamma_{abcd}^n=
    v_c^n (\partial_a F_{bd}^n - \partial_b F_{ad}^n)-\epsilon_{abe}(\partial_c \Omega_e^n) \mathcal M^n_d,
    \label{eq-Gamma}
\end{eqnarray}
where $\epsilon_{abe}$ is the Levi-Civita symbol.

Similar to $\chi^{(0)}$, $\chi^{(1)}$ is also a Fermi surface property, as it should be. From (\ref{15}), it is also a purely dissipationless Hall response, and the response tensor $\chi_{abcd}^{(1)}$ in (\ref{eq-chi1}) is antisymmetric in its first two indices.
Recall that $\chi^{(1)}$ is $\mathcal T$-odd and requires $\mathcal{T}$-symmetry breaking of the original system, e.g., when the system has magnetic ordering or under magnetic proximity effect. One can directly verify that $\chi^{(1)}$ vanishes for a $\mathcal T$-invariant system.
Moreover, as we shall discuss later, $\chi^{(1)}$ exhibits distinct nonzero elements under magnetic crystalline symmetry compared to $\chi^{(0)}$, leading to distinct angular dependence in the response currents with respect to the applied external fields.

\subsection{$\tau^2$-scaling current}
The $O(\tau^2)$ contribution to the nonlinear current is a Drude-like response. Since the nonequilibrium distribution
in (\ref{eq-fdis}) that is order of $\sim \tau^2$ already contains a $E^2$ factor, we should only retain the $E$-independent group velocity of individual electrons in order to obtain $j^{(2)}\sim \tau^2$. The resulting current can be expressed as
\begin{eqnarray}
    \bm j^{(2)}= - \int[\dd{\bm k}]\left( \pdv{\tilde{\varepsilon}_n}{\bm k} \right)
    (\tau \bm E\cdot \partial_{\bm k})^2 f_0(\tilde{\varepsilon}_n),
    \label{eq-tau2E2}
\end{eqnarray}
and we only keep terms that scale as $E^2B$. Here, the $B$ field comes in through the correction to the band energy, as in Eq.~(\ref{11}).

After substituting (\ref{11}) into (\ref{eq-tau2E2}) and collecting the relevant terms, we find that the corresponding conductivity tensor $\chi^{(2)}$  takes the form of
\begin{eqnarray}
    \chi_{abcd}^{(2)}
    &=& -\tau^2
    \int[\dd{\bm k}] \Phi_{abcd}^n(\bm k) f_0'(\varepsilon_n),
    \label{eq-chiabcd2}
\end{eqnarray}
where we have defined
\begin{eqnarray}
    \Phi_{abcd}^n=
    -(\partial_b\partial_a \mathcal M^n_{d})v_c^n
    +(\partial_c\partial_b v_a^n) \mathcal M^n_{d}
    \label{eq-Phi2}.
\end{eqnarray}

$\chi_{abcd}^{(2)}$ recovers the previous theoretical result in Refs.~\cite{ideue_bulk_2017,zhang_theory_2018,he_nonlinear_2019,li_nonreciprocal_2021}. Indeed, this current can be derived within the first-order semiclassical theory, as was done in the previous studies. It is $\mathcal T$-even, as can be verified by the
$\mathcal T$-invariance of Eq.~(\ref{eq-chiabcd2}). Interestingly, by noting that
$v_a^n=\partial_a \varepsilon_n$, one observes that $\chi_{abcd}^{(2)}$ in (\ref{eq-chiabcd2}) is symmetric among its first three indices.
In other words, there is no antisymmetric part in this response, so it does not contain a Hall component.
(For a given driving $E$ field, $\chi^{(2)}$ might give current response transverse to $E$, but it should be interpreted as an anisotropic resistance rather than Hall response.) This feature is in sharp contrast to $j^{(0)}$ and $j^{(1)}$, which are purely Hall responses. Meanwhile, $j^{(2)}$ contributes to the longitudinal transport (i.e., along the driving $E$ field), whereas $j^{(0)}$ and $j^{(1)}$ do not.

Before moving to the model study, we mention that there is not a nonlinear current of order $\tau^3$ in 2D systems. Such a contribution can only arise from the conventional Lorentz force mechanism, which is suppressed in 2D as the cyclotron motion of electrons in $z$ direction is quenched.

\section{Rashba model}

\begin{figure}[tb]
    \centering
    \includegraphics[width=0.48\textwidth]{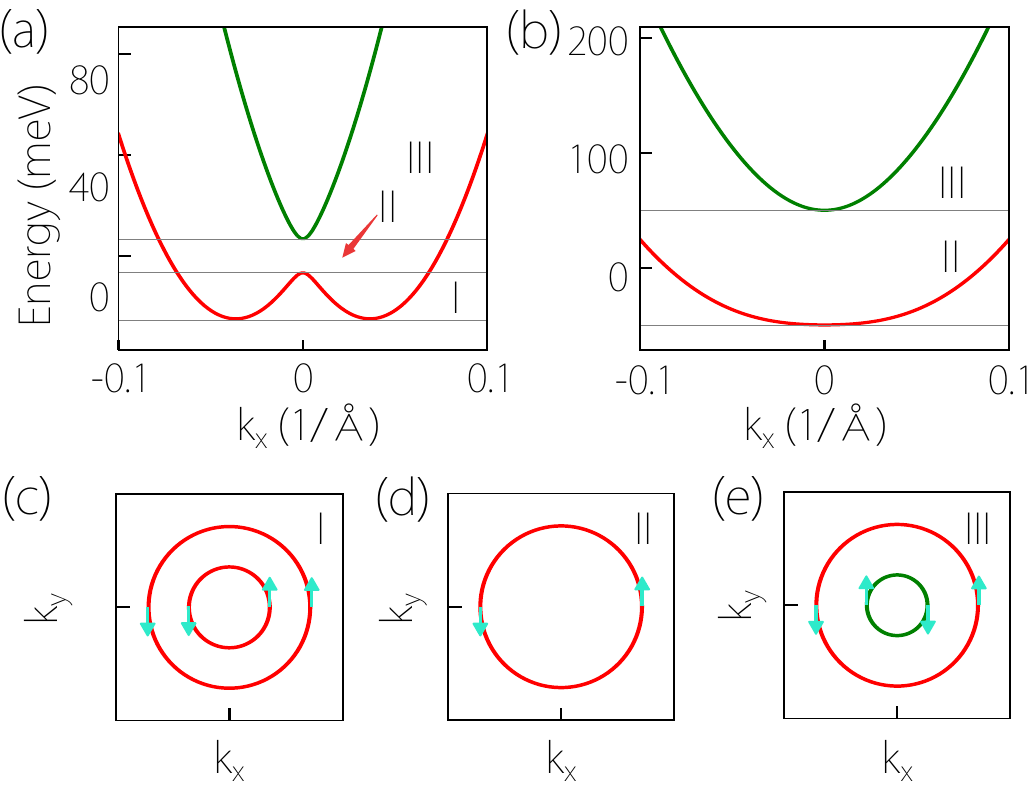}
    \caption{Energy spectrum of Rashba model (\ref{eq-H}) along $k_x$ with $k_y=0$.
    We use the parameters in~\cite{wang_observation_2022}: $m^*=0.28 m_\mathrm{e}$, $g=19.5$, and $\alpha=\SI{1}{eV\cdot\angstrom}$.
    In (a) $\Delta=\SI{5}{meV}$, and in (b) $\Delta=\SI{50}{meV}$, corresponding to the two cases discussed in the main text.
    (c,d,e) show the typical Fermi surface profiles and their in-plane spin structures in three energy regions marked in (a). The green arrows indicate the in-plane spin texture.}
    \label{fig-fig1}
\end{figure}

We shall apply our theory to the Rashba model, which is a widely studied model for spin-orbit coupled 2D electron gases with structural inversion symmetry breaking. Here, we first briefly review the basics of this model.

The Rashba model considered here can be described by the following Hamiltonian:
\begin{eqnarray}
    H= \lambda k^2+ \alpha(k_x \sigma_y - k_y\sigma_x)
    +\Delta \sigma_z
    %+\frac{g \mu_\mathrm{B}B_y}{2}\sigma_y
    \label{eq-H}.
\end{eqnarray}
Here $k=\sqrt{k_x^2+k_y^2}$, $\sigma_i$'s are Pauli matrices representing the spin degree of freedom, $\lambda\sim 1/(2m^*)$ can be related to the effective mass $m^*$ of the system, $\alpha$ is the Rashba SOC strength, and $\Delta$ represents a $\mathcal{T}$-breaking term which appears as an exchange field along $z$ direction and may originate physically from magnetic ordering or magnetic proximity effect~\cite{hauser_magnetic_1969,white_theory_1985}. For convenience, in the following discussion, we assume the model parameters $\lambda$, $\alpha$, and $\Delta$ to be positive unless specified otherwise.

We note the following two points regarding this model. First, due to the $C_{2z}$ symmetry of  this model, all second-order current response, such as those $\sim EB$ or $\sim E^2$, must vanish, whereas our target current $j\sim E^2B$ is allowed.
Second, the presence of the $k$ quadratic term, i.e., the $\lambda$ term in $H$, is necessary for  the current $j\sim E^2B$ to be nonzero. This is because: Without the $\lambda$ term, the Hamiltonian $H$ reduces to the 2D gapped Dirac model, then the in-plane $B$ field, which enters into $H$ via a Zeeman coupling term $g\mu_B\bm B\cdot \bm{\sigma}/2$, can only shift the location of the (gapped) Dirac point but not affect the eigenstates, hence having no effect on the charge current. Thus, the $\lambda$ term must be retained in this model for studying the $j\sim E^2B$ response.

The energy spectrum for this model can be readily solved as $\varepsilon_\pm(\bm k) =  \lambda k^2 \pm  \sqrt{\alpha^2 k^2+\Delta^2}$.
It is well known that there are two different cases regarding the shape of the two bands.
The first is the weak exchange field case $\Delta < \alpha^2/2\lambda$ in which the lower band takes a Mexican hat shape, with band minimum $E_M=-\lambda \Delta^2/ \alpha^2- \alpha^2/4\lambda$ located at a circle with a finite radius $k$ [see Fig.~\ref{fig-fig1}(a)]. Then, the whole spectrum can be divided into three energy regions. Region I is the interval $(E_M,-\Delta)$. For Fermi energy in region I, the Fermi surface consists of two concentric circles, both from the lower band, as shown in Fig.~\ref{fig-fig1}(a). The two circles are said to have the same helicity, meaning that they have the same
kind of spin winding pattern [see Fig.~\ref{fig-fig1}(c)]. Region II corresponds to the interval $(-\Delta,\Delta)$, in which the Fermi surface only consists of a single circle. Finally, region III is given by $(\Delta,\infty)$. The constant energy surface in region III again consists of two circles. However, differing from region I, here, one circle is from the upper band and one from the lower band, and they have opposite helicities, as shown in Fig.~\ref{fig-fig1}(e).

The second is the strong exchange field case $\Delta>\alpha^2/2\lambda$. As illustrated in Fig.~\ref{fig-fig1}(b), in this case, the energy minimum for the lower band occurs at $k=0$ with $E_M=-\Delta$. One can easily see that the key difference from the previous case is that region I disappears, and we only have region II and III [see Fig.~\ref{fig-fig1}(e)].

Since dc transport is a Fermi surface property, we shall see below that the expressions of the nonlinear conductivity tensor differ in the three regions.

\section{Result for Rashba model}
In this section, we apply our theory to the Rashba model in Eq.~(\ref{eq-H}). Below, we first analyze the symmetry properties of the nonlinear conductivity tensor for this model. Then, we will proceed to evaluate the nonzero tensor elements.

\subsection{Symmetry of $\chi$ tensor}\label{SS}
The Rashba model in (\ref{eq-H}) has a quite high symmetry. It is invariant under any rotation along $z$. It also respects the combined $M\mathcal{T}$ symmetry where $M$ is any mirror line in the 2D plane. These symmetries strongly constrain the form of response tensors according to Neumann's principle~\cite{powell_symmetry_2010}. In addition, the three contributions $\chi^{(0)}$, $\chi^{(1)}$, and $\chi^{(2)}$ each has its own symmetry characters. For example, $\chi^{(0)}$ and $\chi^{(1)}$ are antisymmetric in the first two indices, $\chi^{(2)}$ is symmetric among its first three indices, as we have discussed before. Finally, from its definition, only the symmetric component in the middle two indices contributes to the response. We may impose a symmetrization on these two indices and use the standard notation of parenthesis to indicate the symmetrized indices, i.e., $\chi_{a(bc)d}=(\chi_{abcd}+\chi_{acbd})/2$.

Combining all these constraints, we find that for each $\chi^{(i)}$ ($i=0,1,2$), there is only one independent tensor element. Specifically, for $\chi^{(0)}$, the nonzero elements are:
\begin{equation}
  %\chi^{(0)}_{xyxx}=\chi^{(0)}_{xxyx}=-\chi^{(0)}_{yxyy}=-\chi^{(0)}_{yyxy}=\frac{1}{2}\chi^{(0)}_{xyyy}=-\frac{1}{2}\chi^{(0)}_{yxxx}.
    \chi^{(0)}_{x(yx)x}=-\chi^{(0)}_{y(xy)y}=\frac{1}{2}\chi^{(0)}_{xyyy}=-\frac{1}{2}\chi^{(0)}_{yxxx}.
\end{equation}
For $\chi^{(1)}$, these are:
\begin{equation}
  %\chi^{(1)}_{xyxy}=\chi^{(1)}_{xxyy}=\chi^{(1)}_{yxyx}=\chi^{(1)}_{yyxx}=-\frac{1}{2}\chi^{(1)}_{xyyx}=-\frac{1}{2}\chi^{(1)}_{yxxy}.
    \chi^{(1)}_{x(yx)y}=\chi^{(1)}_{y(xy)x}=-\frac{1}{2}\chi^{(1)}_{xyyx}=-\frac{1}{2}\chi^{(1)}_{yxxy}.
\end{equation}
Finally, for $\chi^{(2)}$, the nonzero elements are
\begin{align}
    &\chi_{xyyy}^{(2)}=\chi_{y(xy)y}^{(2)}=
    -\chi_{yxxx}^{(2)}
    \nonumber\\
    &=-\chi_{x(xy)x}^{(2)}=
    \frac{\chi_{xxxy}^{(2)}}{3}=-\frac{\chi_{yyyx}^{(2)}}{3}.
\end{align}

\begin{figure}[tb!]
    \centering
    \includegraphics[width=0.45\textwidth]{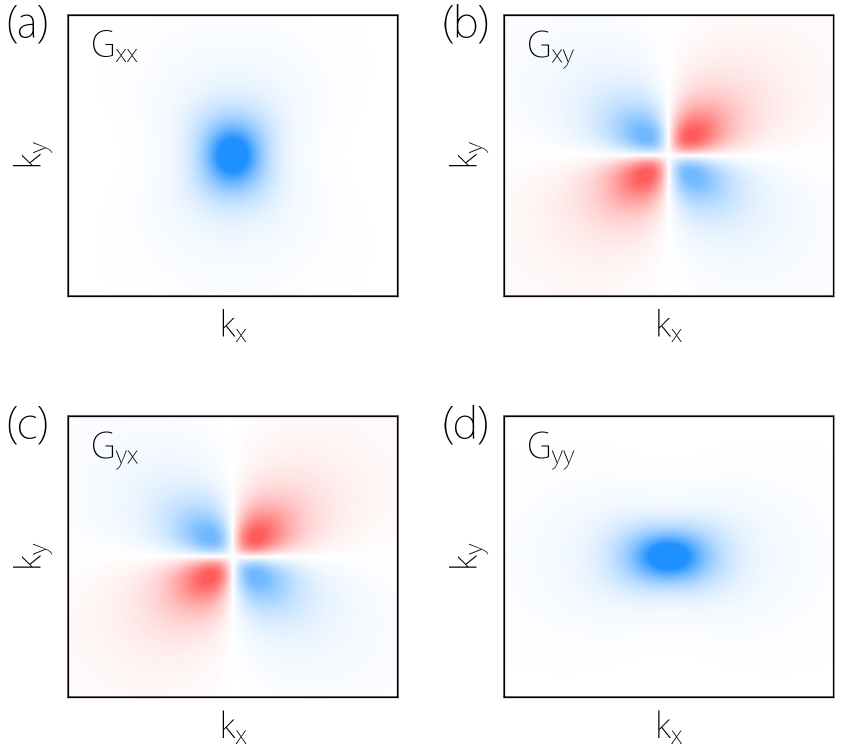}
    \caption{Distribution of the BCP tensor components in $k$-space for the lower Rashba band. The center of each figure is at $k=0$. Model parameters are the same as in Fig.~\ref{fig-fig1}(a).}
    \label{fig-Gtensor}
\end{figure}

\subsection{$\tau^0$-scaling conductivity}

The expression of $\chi^{(0)}$ involves two geometric quantities $G$ (BCP) and $\Lambda$ (spin susceptibility of BCP). Let us first take a look at these tensors for the Rashba model~\eqref{eq-H}.

\begin{figure}[tb]
    \centering
    \includegraphics[width=0.45\textwidth]{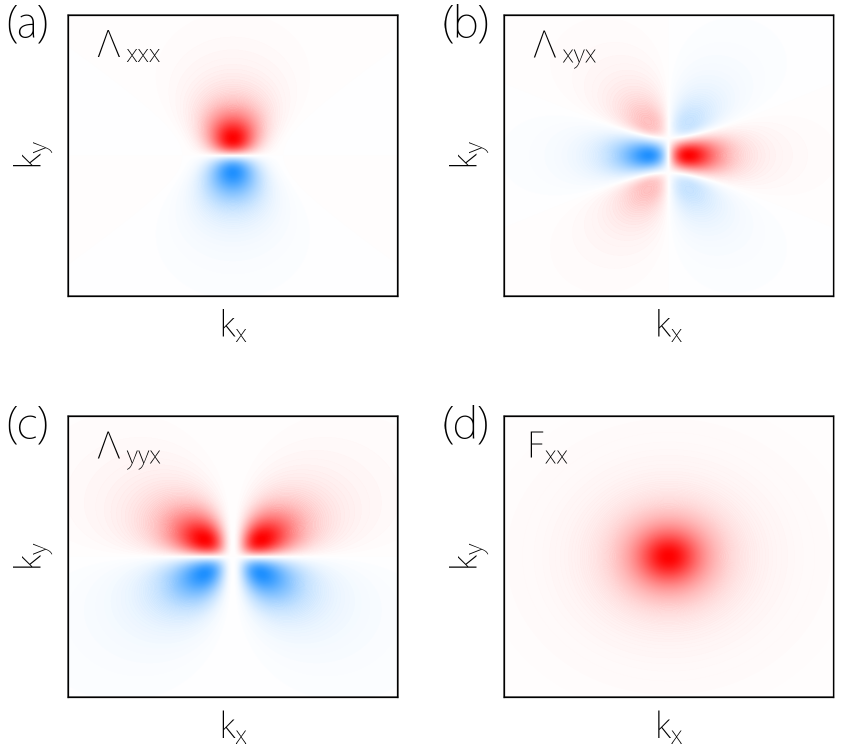}
    \caption{Distributions of the spin susceptibility of BCP tensor components (a) $\Lambda_{xyx}$, (b) $\Lambda_{xyx}$, (c) $\Lambda_{yyx}$, and of the ASP component $F_{xx}$, for the lower Rashba band. The
    parameters are the same as in Fig.~\ref{fig-fig1}(a).}
    \label{fig-Ltensor}
\end{figure}

For example, considering the lower band, the BCP tensor $G$ can be calculated as
\begin{eqnarray}
    [G_{ab}]=  \frac{ \alpha^4}{4\zeta^5}\mqty[-k_y^2-\Delta^2/\alpha^2 & k_xk_y \\  k_xk_y & -k_x^2-\Delta^2/\alpha^2]
    \label{eq-Gab},
\end{eqnarray}
where $\zeta=\sqrt{\alpha^2 k^2+\Delta^2}$. In Fig.~\ref{fig-Gtensor}, we plot the distribution of the $G$ tensor elements in $k$ space. One observes that the diagonal elements $G_{xx}$ and $G_{yy}$ take a monopole shape, whereas the off-diagonal element $G_{xy}$ ($G$ is by definition a symmetric tensor) has a quadrupole like distribution. All elements are mainly concentrated in a region around $k=0$, where the local gap between the two bands is small. This is a common feature that is also observed in previous studies \cite{lai_third-order_2021,liu_intrinsic_2021,liu_berry_2022}.

In Fig.~\ref{fig-Ltensor}(a,b,c), we plot the $k$-space distribution of some $\Lambda$ tensor elements that are involved in $\chi^{(0)}$. One observes that although the pattern may be more complicated (e.g., $\Lambda_{xyx}$ has a hexapole type pattern), the feature that these quantities are concentrated around the small gap region (around $k=0$) remains the same, which is typical for band geometric quantities encoding the information of interband coherence.

\begin{table*}
        \setlength\extrarowheight{8pt}
    \centering
    \begin{tabular}{p{0.1\textwidth}<\centering p{0.3\textwidth}<\centering p{0.3\textwidth}<\centering p{0.3\textwidth}<\centering }
        \hline\hline
        & I & II & III
        \\ \hline
        $\chi^{(0)}_{xyyy}$ &
        $\frac{e^3  g\mu_\mathrm{B}}{8\pi\hbar} \frac{ (2\lambda \mu+ \alpha^2)}{\mu^2  \alpha^2} \eta^4 \sqrt{\frac{\lambda}{\mu-E_R}}$ &
        $\frac{e^3 g\mu_\mathrm{B}}{\pi\hbar}
        \frac{\lambda^{5/2}(2\lambda \mu+ \alpha^2+2 \alpha\sqrt{\lambda(\mu-E_R)})}{ \alpha^2\sqrt{\mu-E_R}( \alpha+2\sqrt{\lambda(\mu-E_R)})^4}$ &
        $-\frac{e^3 g\mu_\mathrm{B}}{4\pi\hbar} \frac{\lambda \eta^4}{\mu^2  \alpha}$
        \\
        $\chi^{(1)}_{xyyx}$ &
        $- \frac{e^3  \tau g \mu_\mathrm{B}}{8\pi \hbar^2} \frac{3  \Delta (2\lambda \mu+ \alpha^2)}{\mu^2  \alpha^2} \eta^4 \sqrt{\frac{\lambda}{\mu-E_R}}$ &
        $\frac{e^3 \tau g\mu_\mathrm{B}}{\pi\hbar^2} \frac{3 \Delta\lambda^{5/2}(2\lambda \mu+ \alpha^2+2 \alpha\sqrt{\lambda(\mu-E_R)})}{ \alpha^2\sqrt{\mu-E_R}( \alpha+2\sqrt{\lambda(\mu-E_R)})^4}$ &
        $\frac{ 3e^3 \tau g \mu_\mathrm{B}}{4\pi \hbar^2} \frac{ \lambda \Delta \eta^4}{\mu^2  \alpha}$
        \\
        $\chi^{(2)}_{xyyy}$ &
        $ \frac{e^3 \tau^2 g\mu_\mathrm{B}}{8\pi \hbar^3} \eta^4 \sqrt{ \frac{\lambda}{\mu-E_R}} $ &
        $\approx \frac{e^3 \tau^2 g \mu_\mathrm{B}}{8\pi \hbar^3} \frac{\lambda}{\sqrt{4 \lambda \mu+ \alpha^2}} $ &
        $\approx -\frac{e^3 \tau^2 g\mu_\mathrm{B}}{4\pi \hbar^3} \frac{\lambda \Delta^2 \eta^4 }{\mu^2  \alpha} $
        \\ \hline \hline
    \end{tabular}
    \caption{Results of the three contributions to $\chi$ in different energy regions for the Rashba model. Here, we define a dimensionless parameter $\eta=\mu  \alpha^2/(\lambda\Delta^2+\mu  \alpha^2)$. The expressions for $\chi^{(0)}_{xyyy}$ and $\chi^{(1)}_{xyyx}$ in Regions II and III apply to both the strong and weak exchange field cases, whereas those for $\chi^{(2)}_{xyyy}$ are approximate results valid only for the weak exchange field case.
}
    \label{tab:taun}
\end{table*}

Based on Eqs.~(\ref{eq-chiabcd0}, \ref{lambda}), we obtain analytic expressions of $\chi^{(0)}$ for the Rashba model. Since there is only one independent element for this model, we choose to consider  $\chi^{(0)}_{xyyy}$.
The results are listed in Table~\ref{tab:taun}. Here, we restore the factors of $e$ and $\hbar$ in the expressions.  Note that the expression for the nonlinear conductivity differs for Fermi level $\mu$ lying in the three different energy regions, due to the different Fermi surface characters. It is worth noting that the results vanish when $\lambda=0$, confirming our previous observation that the quadratic term in Rashba model is needed for a nonzero signal.

\begin{figure}[tb]
    \centering
    \includegraphics[width=0.48\textwidth]{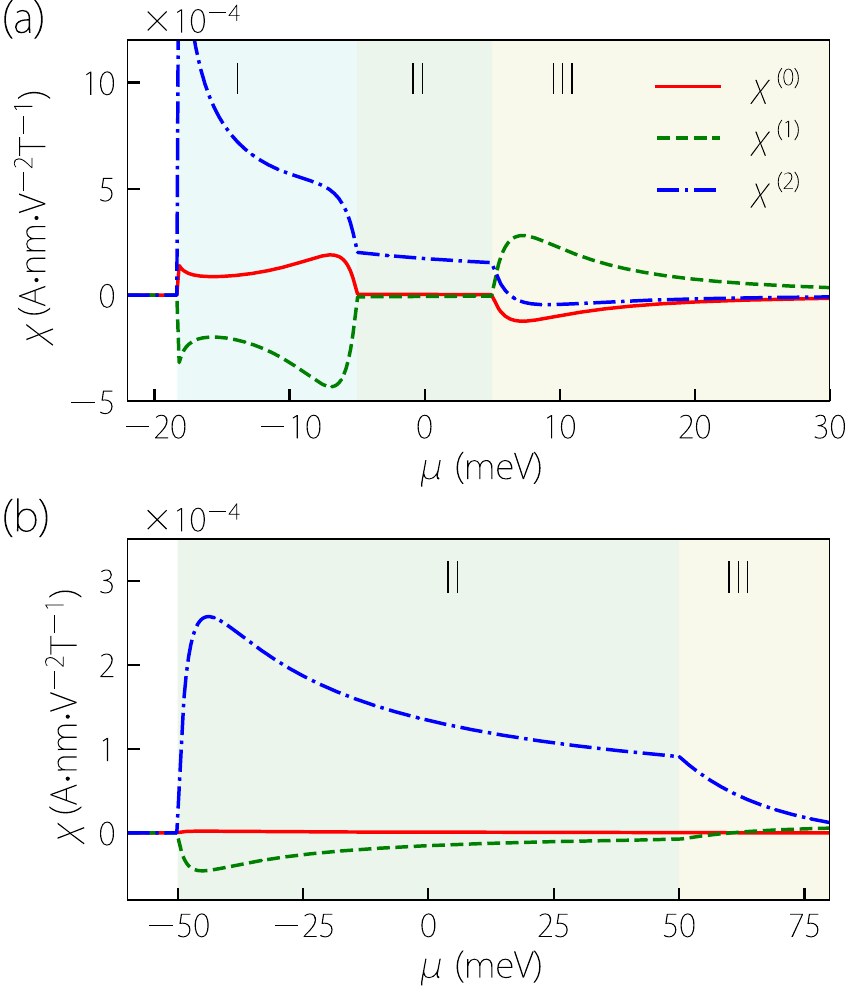}
    \caption{(a) The chemical potential dependence of the three contributions to $\chi$. The two figures correspond
        to the two cases in Fig.~\ref{fig-fig1}(a) and (b), respectively.
    Here, we take $\tau=\SI{100}{fs}$, and other model parameters are the same as in Fig.~\ref{fig-fig1}. }
    \label{fig-chimu}
\end{figure}

In Fig.~\ref{fig-chimu}(a), we plot the numerical result of $\chi^{(0)}_{xyyy}$ as a function of $\mu$ (red solid line) for the case with relatively small $\Delta$. One observes the following features. First, the value of $\chi^{(0)}$ is sizable in region I and III, and it is peaked around the band edge energies corresponding to the small gap. Here, the small local (direct) gap is at $k=0$, and the two band edges are at energies $\pm \Delta$. This can be understood from the behavior of the geometric quantities $G$ and $\Lambda$ discussed above. Second, the two peaks have opposite values. This can be readily seen from the analytic formulas in Table~\ref{tab:taun}. Finally, $\chi^{(0)}$ is much smaller (but nonzero) in region II. In this region, there is only a single Fermi circle, on which the local gap is quite large. The large local gap suppresses the $G$ and $\Lambda$ tensors, hence $\chi^{(0)}$ is small.

We also consider the case with large $\Delta$, and the result is shown in Fig.~\ref{fig-chimu}(b). As discussed, in this case, there are only regions II and III. The amplitude in this case is rather small due to the large local energy gaps.
%The qualitative behavior is similar to Fig. for the corresponding region. ...

\subsection{$\tau^1$-scaling conductivity}

The formula for $\chi^{(1)}$ involves the ASP tensor $F_{ab}$. For the lower band of the Rashba model, direct calculation gives the diagonal element
of this tensor as
\begin{equation}
  F_{xx}=F_{yy}= {g\mu_\mathrm{B} \alpha \Delta}/{\zeta^3},\label{27}
\end{equation}
and the off-diagonal elements are zero. Again, this tensor is peaked around the small gap at $k=0$ [see Fig.~\ref{fig-Ltensor}(d)], as can be easily observed from its $\zeta^{-3}$ dependence. The appearance of $\alpha$ and $\Delta$ in the numerator of (\ref{27}) indicates that both SOC and exchange term are needed for a nonzero ASP.

Analytic expressions for $\chi^{(1)}_{xyyx}$ in the three energy regions are shown in Table~\ref{tab:taun}. One observes that besides a finite $\lambda$, a nonzero $\chi^{(1)}$ must require a finite $\Delta$. This verifies our previous analysis that the time-reversal symmetry breaking is necessary for the $\chi^{(1)}$ response.

In Fig.~\ref{fig-chimu}(a), we plot the numerical results for $\chi^{(1)}_{xyyx}$ as a function of $\mu$. One can see that except for an opposite sign, its qualitative behavior is quite similar to  $\chi^{(0)}$. Namely, it exhibits peaks around the small gap edges in region I and III, and it is suppressed in region II. This is expected, because both  $\chi^{(0)}$ and $\chi^{(1)}$ depend on band geometric quantities that are peaked around the small gap edges. In fact, from Table~\ref{tab:taun}, we find that for Rashba model, there is a simple relation between  $\chi^{(1)}$ and $\chi^{(0)}$, namely,
\begin{equation}
  \frac{\chi_{xyyx}^{(1)}}{\chi_{xyyy}^{(0)}}= -3\frac{\Delta}{\hbar /\tau}.
\end{equation}

\subsection{$\tau^2$-scaling conductivity}

The analytic results of $\chi^{(2)}$ are summarized in Table~\ref{tab:taun}.
Differing from $\chi^{(0)}$ and $\chi^{(1)}$, as a Drude-like conductivity, $\chi^{(2)}$ is not related to geometric quantities that are enhanced at small gaps, which can already be seen from formulas (\ref{eq-chiabcd2}) and (\ref{eq-Phi2}).
As a result, the $\chi^{(2)}$  versus $\mu$ curve exhibits behaviors distinct from $\chi^{(0)}$ and $\chi^{(1)}$.

From the numerical result in Fig.~\ref{fig-chimu}(a), one can observe the following features regarding $\chi^{(2)}$. First, unlike $\chi^{(0)}$ and $\chi^{(1)}$, it does not show peaks at the small gap edges. Second, in region I, $\chi^{(2)}$'s value increases as $\mu$ approaches the band minimum $E_M$, connected with the large density of states there. Third, its value is still sizable in region II. Since $\chi^{(0)}$ and $\chi^{(1)}$ are suppressed in this region, $\chi^{(2)}$ becomes the dominant contribution there. Finally, its value is small in region III. We find that the contributions from the two Fermi circles in this region tend to cancel each other due to their opposite helicities. In fact, one can observe from Table~\ref{tab:taun} that the leading order of $\chi^{(2)}\propto \Delta^2$ in region III, so it would vanish identically when $\Delta=0$.

\section{Discussion and Conclusion}

From the above model study, we can learn some general features of the nonlinear conductivity $\chi$. First, the Hall responses  $\chi^{(0)}$ and $\chi^{(1)}$ exhibit `band geometric' characters, i.e., they are strongly enhanced at edges of small local gaps. This is a manifestation of their origin from interband coherence effects. In comparison, $\chi^{(2)}$ is a more `classical' Drude-like contribution. It has more dependence on carrier density and could dominate in `trivial' energy windows where local gaps around Fermi level are large.

\begin{figure}[tb]
    \centering
    \includegraphics[width=0.42\textwidth]{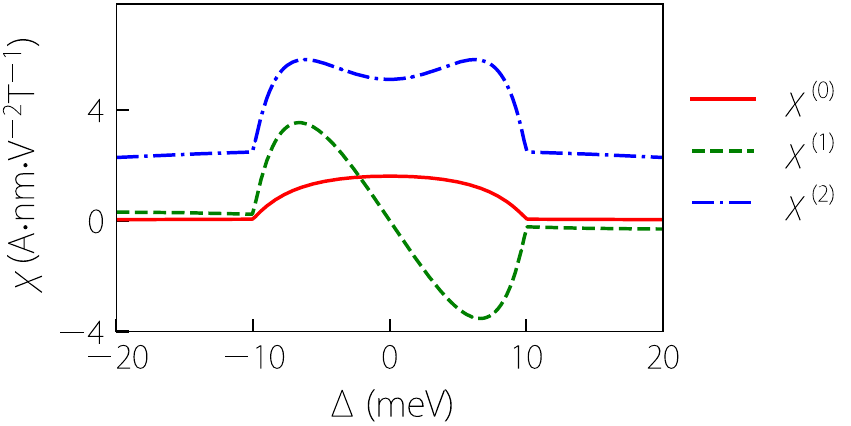}
    \caption{Dependence of the three contributions to $\chi$ on $\Delta$. Here, we take $\tau=\SI{100}{fs}$ and $\mu=\SI{-10}{meV}$, and other model parameters are the same as in Fig.~\ref{fig-fig1}(a).}
    \label{fig-chiDelta}
\end{figure}

In analyzing experimental data, a commonly used method is to plot nonlinear conductivity (or other measured quantities) against the linear longitudinal conductivity $\sigma$ which is linear in $\tau$ (e.g., by varying temperature or other system control parameters). This helps to separate contributions that have different scaling dependence on $\tau$~\cite{lai_third-order_2021,kang_nonlinear_2019,wang_room-temperature_2022}. It can certainly be applied here to separate out the $\chi^{(i)}$'s $(i=0,1,2)$, which have been grouped according to their $\tau$ scaling.

In addition, the different symmetry characters of the $\chi^{(i)}$'s may also be utilized for their separation. For example, as we mentioned, $\chi^{(1)}$ must require $\mathcal T$ breaking, whereas $\chi^{(0)}$ and $\chi^{(2)}$ do not. From the model results in Table~\ref{tab:taun}, one can immediately observe that $\chi^{(0)}$ and $\chi^{(2)}$ are even functions of the exchange field $\Delta$, whereas  $\chi^{(1)}$ is an odd function. This behavior is also illustrated in Fig.~\ref{fig-chiDelta}, which is consistent with the symmetry character of $\mathcal T$-even and $\mathcal T$-odd. Using this property, one can separate $\chi^{(1)}$ from $\chi^{(0)}$ and $\chi^{(2)}$  by analyzing the response under the reversal of magnetic ordering.

\begin{figure}[tb]
    \centering
    \includegraphics[width=0.48\textwidth]{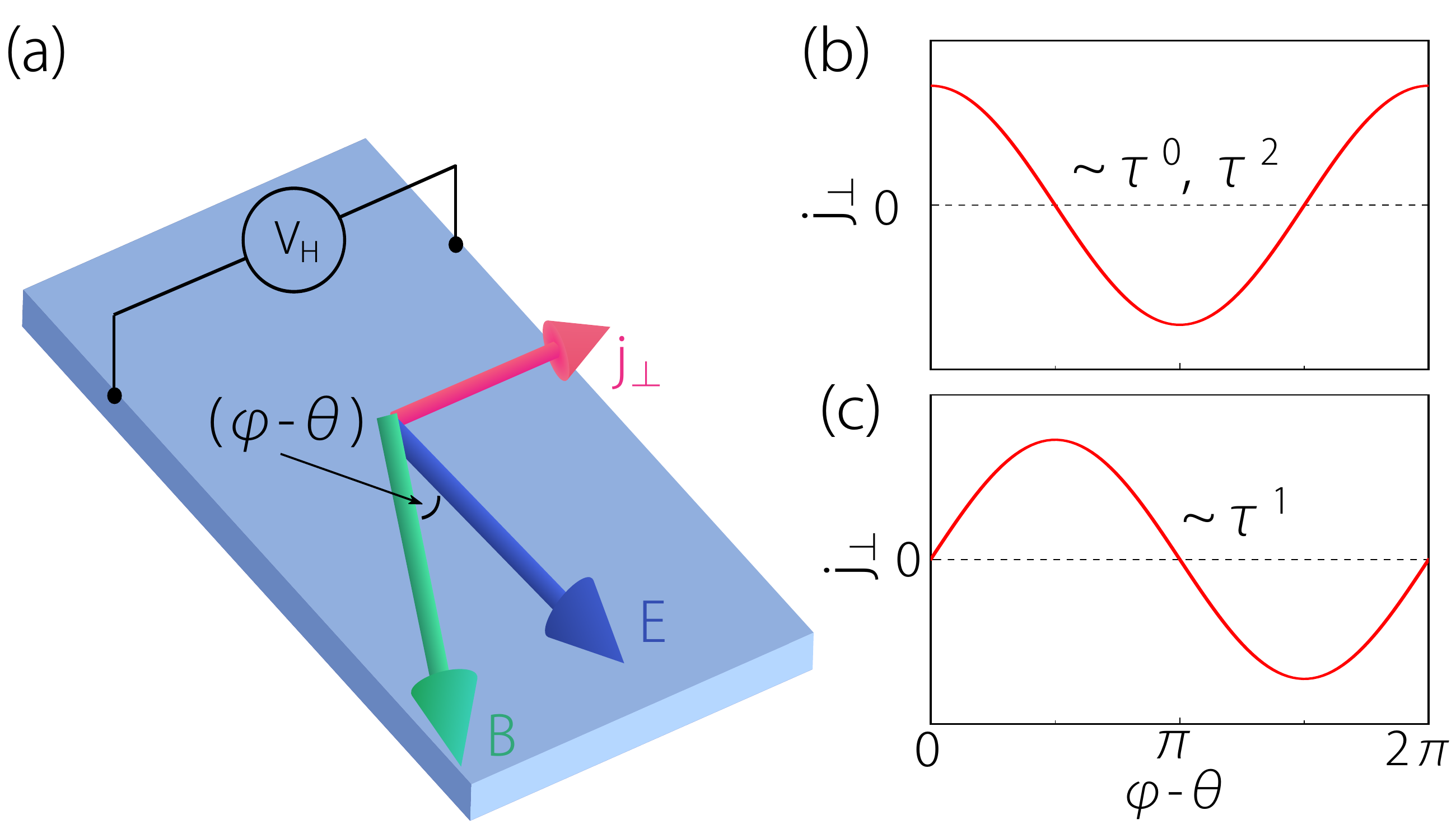}
    \caption{(a) Schematic measurement setup for the nonlinear planar Hall effect. (b) The transverse currents from $\tau^0$- and $\tau^2$-scaling contributions in
    the Rashba model are cosine functions of the relative angle between $E$ and $B$ fields, (c) whereas the $\tau^1$-scaling contribution
    is  a sine function.}
    \label{fig-fig}
\end{figure}

Another important difference in symmetry characters is regarding the directions of the driving field and the response current. We have shown that $\chi^{(0)}$ and $\chi^{(1)}$ are purely Hall responses, whereas $\chi^{(2)}$ contains both longitudinal and transverse responses but does not have a Hall component. Moreover, as we have seen in the model study, crystalline symmetries usually result in different nonzero elements for $\chi^{(i)}$'s. This will lead to their different angular dependence when varying the field/current direction.

To demonstrate this point, we use the Rashba model as an example. In practice, one usually defines the coordinate axis to be along some crystal axis direction. The applied $E$ and $B$ fields are not necessarily aligned with such a direction. In fact, a commonly used technique is to fabricate multiple leads at different angles on a disk shaped sample, so that one can vary the direction of driving $E$ field and measure the response at any other direction.  Assume the applied $E$ and $B$ field directions are specified by
$\bm E=E(\cos\theta,\sin\theta)$ and $\bm B=B(\cos\varphi,\sin\varphi)$. Now, using the symmetry character discussed in Sec.~\ref{SS}, we find that $j^{(0)}$ flows only in the transverse direction, with $\bm j^{(0)}=j^{(0)}(-\sin\theta,\cos\theta)$, and the magnitude exhibits the following angular dependence:
\begin{equation}
  j^{(0)}/(E^2B)=\chi^{(0)}_{xyyy}\cos(\theta-\varphi),
\end{equation}
which has a cosine dependence on the relative angle between $E$ and $B$ fields. Therefore, the response is maximal when the two fields are aligned and vanishes when they are perpendicular. As for $j^{(1)}$, it is also in the transverse direction, but its angular dependence is different:
\begin{equation}
  j^{(1)}/(E^2B)=\chi^{(1)}_{xyyx}\sin(\theta-\varphi),
\end{equation}
which becomes maximal when the fields are orthogonal and vanishes when they are aligned. Meanwhile, $j^{(2)}$ has both transverse component $j^{(2)}_\bot$ and longitudinal component $j^{(2)}_\|$. They show the following angular dependence:
\begin{eqnarray}
    j_\bot^{(2)}/(E^2B)&=&  \chi_{xyyy}\cos(\theta-\varphi)
    \label{eq-jbot2},
    \\
    j_\|^{(2)}/(E^2B)&=&  3\chi_{xyyy}\sin(\theta-\varphi).
    \label{eq-j||2}
\end{eqnarray}
The analysis above is done for the Rashba model. For a given material, the constrains will come from its specific crystalline symmetry. The generally different angular dependence not only helps to separate the different contributions, but also offers a route to control the nonlinear signal which may be useful for nonlinear device design.

Finally, we remind that the presented theory is within the relaxation time approximation, which neglects some possible delicate disorder-induced effects. Prominent examples of such effects include side jump and skew scattering of linear and nonlinear electrical Hall effects~\cite{nagaosa_anomalous_2010,konig_gyrotropic_2019,du_disorder-induced_2019,xiao_theory_2019,nandy_symmetry_2019,isobe_high-frequency_2020}. Similar effects have not been investigated in the context of planar Hall effect yet, and more systematic studies are needed for future research. It is noted that these disorder induced effects are highly sensitive to details of disorder contents, which are usually hard to determine in practice. In this regard, our present theory has the advantage to be readily amenable to quantitative evaluation (e.g., using first-principles calculations) for real materials and serves as a benchmark for comparing theory and experiment.

In conclusion, we have presented a unified theory for the nonlinear charge current response under an in-plane magnetic field in 2D spin-orbit coupled systems, which not only includes previously proposed $\mathcal{T}$-even contribution but also puts forward a new $\mathcal{T}$-odd contribution of band geometric origin in 2D magnetic systems. We demonstrated that the novel quantum geometric quantity ASP plays an important role in this $\mathcal{T}$-odd nonlinear planar Hall effect. The different symmetry characters of $\mathcal{T}$-even and $\mathcal{T}$-odd nonlinear current responses in magnetic systems lead to distinctive angular dependence with respect to applied fields and render a powerful tool for distinguishing different contributions in experiment. Our theory is demonstrated by studying the 2D spin-polarized Rashba model. We also highlight that the nonlinear currents of band geometric origin, including the intrinsic and the $\mathcal{T}$-odd one, are enhanced significantly by band near-degeneracies around Fermi surfaces, and they may dominate for systems hosting topological band features.
%Moreover, the angular-dependence may offer a route to control the nonlinear response and to switch it on and off by simply rotating the $B$ (or $E$) fields.
%This could be useful for the design of novel nonlinear rectification or frequency doubling devices.

\begin{acknowledgements}
    The authors thank D. L. Deng for helpful discussions. This work was supported by
    the Research Grants Council of Hong Kong (Grants No. CityU 21304720, CityU 11300421, and C7012-21G),
    City University of Hong Kong (Project No. 9610428),
    UGC/RGC of Hong Kong SAR (AoE/P-701/20),
    and Singapore NRF CRP22-2019-0061.
\end{acknowledgements}

\bibliographystyle{apsrev4-2}
%\bibliography{rashba_ref}
\bibliography{ref}

\end{document}